# Flux line lattice structure and behaviour in anti-phase boundary free vicinal YBa$_2$Cu$_3$O$_{7-\delta}$ thin films

J.H. Durrell[a)], S. H. Mennema[a,b)], Ch. Jooss[c)], G. Gibson[a)], Z.H. Barber[a)], H.W. Zandbergen[b)] and J.E. Evetts[a)].

[a)] Department of Materials Science and Metallurgy, University of Cambridge, Pembroke Street, Cambridge, CB2 3QZ, UK.

[b)] National Centre for HREM, Delft University of Technology, Rotterdamseweg 137, 2628 AL Delft, The Netherlands.

[c)] Institut für Materialphysik, Windausweg 2, D-37073 Göttingen, Federal Republic of Germany.

Field angle dependent critical current, magneto-optical microscopy, and high resolution electron microscopy studies have been performed on YBa$_2$Cu$_3$O$_{7-\delta}$ thin films grown on mis-cut substrates. High resolution electron microscopy images show that the films studied exhibited clean epitaxial growth with a low density of anti-phase boundaries and stacking faults. Any anti-phase boundaries (APBs) formed near the film substrate interface rapidly healed rather than extending through the thickness of the film. Unlike vicinal films grown on annealed substrates, which contain a high density of anti-phase boundaries, magneto optical imaging showed no filamentary flux penetration in the films studied. The flux penetration is, however, asymmetric. This is associated with intrinsic pinning of flux strings by the tilted *a-b* planes and the dependence of the pinning force on the angle between the local field and the *a-b* planes. Field angle dependent critical current measurements exhibited the striking vortex channelling effect previously reported in vicinal films. By combining the results of three complementary characterisation techniques it is shown that extended APB free films exhibit markedly different critical current behaviour as compared to APB rich films. This is attributed to the role of APB sites as strong pinning centres for Josephson string vortices between the *a-b* planes.

INTRODUCTION

Measurements on thin superconducting films grown on vicinal substrates allow effects due to the experimental geometry to be separated from those due to the orientation of the applied magnetic field with respect to the crystallographic axes of the superconductor. [1,2] One example of such an effect is the vortex channeling effect observed in vicinal $YBa_2Cu_3O_{7-\delta}$ films when a magnetic field is applied parallel to the *a-b* plane direction. [3,4] These films exhibit a critical current density ($j_c$) parallel to the vicinal steps similar to that found in *c*-axis films and a suppressed $j_c$ for currents flowing perpendicular to the steps where *c*-axis directed current transport must occur.

It is possible to produce YBCO thin films grown on mis-cut (vicinal) substrates which exhibit a high density of anti-phase boundaries (APBs). [5,6] In contrast to the critical current behaviour observed in microstructurally clean films these films exhibit enhanced critical current behaviour in the *a-b* plane direction as compared to *c*-axis oriented films. [7,8] This enhancement has been attributed to pinning due to anti-phase boundaries in the film. Additionally films grown on mis-cut $LaAlO_3$ substrates have been shown to grow with a high density of defects that act as strong pinning centres [9]. In this paper the results of angular critical current, magneto-optical imaging (MO), and transmission electron microscopy studies are reported from thin films of $YBa_2Cu_3O_{7-\delta}$ grown on vicinal substrates with a very low density of APB and stacking fault (SF) defects.

A series of 220 nm thick thin films of $YBa_2Cu_3O_{7-\delta}$ were prepared on vicinal (100) $SrTiO_3$ substrates with varying mis-cut angles, the films were deposited by on-axis pulsed laser deposition. The substrates were not annealed before deposition. The deposition process is identical to that used for the preparation of *c*-axis oriented $YBa_2Cu_3O_{7-\delta}$ films except that that the best quality films are grown at a slightly higher temperature. X-Ray diffraction studies confirmed that epitaxial growth had occurred on the vicinal substrate. Fig. 1 shows the geometry of such a vicinal film defining the orientation ($\theta,\phi$) of an applied magnetic film with respect to the crystal structure, the vicinal tilt angle $\theta_v$ between the *c*-axis and a vector **n** normal to the surface of the film. Two directions may be defined in the plane of the film, **L**, parallel to the *a-b* planes and, **T**, transverse to the vicinal steps.

TEM specimens were prepared by cutting along the **T** direction of the film. A Philips CM30UT transmission electron microscope equipped with a field emission gun was operated at 300 kV to investigate the microstructure of the film.

The response of the films to an applied magnetic field in the absence of a transport current was studied using magneto-optical imaging. The normal component of the magnetic flux density $B_z(x,y)$ of the films is determined using an iron garnet indicator layer [10] as a field sensing element. The measured light intensity distribution is converted to a magnetic flux density distribution by applying a non-linear calibration function [11], taking the intrinsic properties of the iron garnet into account. The spatial resolution of the measurement is 6 μm and the magnetic sensitivity is 0.1 mT. For the determination of the current density distribution, we use a model-independent method, based on a fast two dimensional numerical inversion of Biot-Savart's law [12]. It allows the determination of the in-plane current densities $j_x(x,y)$ and $j_y(x,y)$ with almost the same spatial resolution as the measured $B_z(x,y)$.

Transport critical current versus field angle measurements were performed using a two-axis goniometer mounted in a 8 T cryostat [13]. The films were patterned using optical lithography and Ar-ion milling to provide 10 x 100 μm current tracks oriented both in the **L** direction parallel to the *a-b* planes and in the **T** direction across the vicinal steps. Critical current values were obtained from four terminal current versus voltage measurements on the microbridges using a voltage criterion of $5 \times 10^{-7}$ V.

FILM MICROSTRUCTURE

$YBa_2Cu_3O_{7-\delta}$ films grown by physical vapour deposition on UHV annealed vicinal substrates exhibit a microstructure rich in both APB and SF type defects [8]. Due to its crystallographic mis-cut, the vicinal substrate surface will typically exhibit a terraced washboard structure. The offset between adjacent terraces is an integral multiple of the unit cell height of $SrTiO_3$ (0.39 nm) which, except for particular mis-cut angles, does not correspond to the height of the $YBa_2Cu_3O_{7-\delta}$ unit cell. When growing, $YBa_2Cu_3O_{7-\delta}$ domains coalesce on adjacent terraces and anti-phase boundaries (APBs) with a planar structure are formed. These APBs are extended along the *c*-axis over some unit cells of $YBa_2Cu_3O_{7-\delta}$, as shown in Fig. 2.

A typical microstructure of the $YBa_2Cu_3O_{7-\delta}$ thin films on $SrTiO_3$ vicinal substrates used for this study is shown in Fig. 3. The region above the interface is slightly distorted due to the substrate surface step structure, SFs and short APBs are commonly observed defects in this region. However, APBs are healed after a few unit cells from the interface by inclusion of a SF lying in the (001) plane, as shown in the inset to Fig. 3. Alternatively, the formation of APBs can be prevented by a variable stacking mode of the YBCO on the adjacent terraces. A change of the stacking sequence on adjacent terraces allows a smooth and defect-free overgrowth of a substrate step. The variable stacking mode reduces the number of APBs initiated from the surface steps and permits defect-free matching of the following upper layers, as has also been reported by Pedarnig *et al* [14]. As a consequence of these mechanisms, after the first ~10 nm of film growth a regular microstructure is obtained. In the remaining thickness of the film, the microstructure continues to exhibit this undisturbed character and no growth mechanism related defects are observed.

CRITICAL CURRENT MEASUREMENTS AND VORTEX CHANNELING

In contrast to the majority of high-$T_c$ cuprate superconductors $YBa_2Cu_3O_{7-\delta}$ has a relatively small anisotropy parameter ($\gamma$~5-7) [15] . As a consequence of this the flux line lattice (FLL) exhibits a transition as the angle $\theta$ between the applied magnetic field and the cuprate planes is reduced. For large $\theta > \theta_1$ a lattice of rectilinear but elliptically cored Abrikosov vortices is expected, below $\theta_1$ the flux lines become distorted, whilst for small $\theta < \theta_2$ a fully kinked lattice of flux lines consisting of vortex pancake and Josephson string elements is predicted [15]. The values of $\theta_1$ is given by

$$\tan(\theta_1) = d/\xi_{ab}(t) \tag{1}$$

and that of $\theta_2$ by

$$\tan(\theta_2) = \varepsilon \tag{2}$$

where $t$ is the reduced temperature $T/T_c$ and $\varepsilon$ is the Ginzburg-Landau anisotropy parameter. At $t=0$, $\theta_1=35°$ and $\theta_2=11°$. Above a crossover critical temperature $T_{cr}$~80 K the *c*-axis coherence length is longer than the interlayer spacing and the FLL is always formed of conventional Abrikosov vortices. As a consequence of this field angle dependent crossover no single critical current, $j_c(\theta)$, or pinning force density, $F_L(\theta)$, scaling law will apply to all angles of applied field.

For the kinked vortex structure the pancake pinning and the string pinning contributions are quite different [16]. The pinning force on vortex string elements is generally lower than that on the vortex pancakes, this accounts for the striking vortex channelling effect observed in vicinal films for currents in **T** tracks when the field is applied at angles close to θ=0° [3]. In contrast to the case of *c*-axis films, in a 'vicinal' film this orientation leads to a component of the Lorentz force pushing the string elements along the cuprate chain layers in the weakly pinned direction.

Fig. 4 shows the results of critical current measurements on a current track patterned in a 4° 'vicinal' $YBa_2Cu_3O_{7-\delta}$ film. $j_c(\theta)$ characteristics were obtained at ϕ=0° for a range of temperatures. At θ=0° the channelling minimum is clearly seen at temperatures below $T_{cr}$=80 K. Visible at θ~4° is the peak in the critical current associated with the force free regime. In a *c*-axis film these two peaks are superposed and the channelling minimum is not seen.

Fig. 5 shows the variation in critical current at θ=90° with the vicinal tilt angle, this shows the expected behaviour that as $\theta_v$ increases the critical current measured in the **T** direction decreases. This is consistent with the increasing importance of c-axis transport as $\theta_v$ increases.

Vortex channelling means that the pinning force density for flux string motion inside the blocking layers is smaller than that for the vortex pancakes. APBs would be expected to strongly pin vortex strings inside the blocking layers as they are extended correlated defects. This implies that the greatly enhanced $J_c$ seen in vicinal oriented APB rich films is due to the fact that string elements are no longer weakly pinned for motion inside the blocking layers.

MAGNETO-OPTICAL IMAGING

The anisotropic crystal structure of $YBa_2Cu_3O_{7-\delta}$ leads to a different current distribution in films with macroscopically tilted $CuO_2$ planes as compared to *c*-axis oriented films. The critical current density in the *a-b* planes is larger than that crossing the planes in the *c*-axis direction [17,18]. As a consequence of this, due to the boundary condition that $j(x,y)$ has to flow parallel to the superconductor surface, a macroscopically tilted *a-b* plane immediately leads to anisotropy of the critical current density, where critical current in the **T** direction, $j_{c,T}$, is smaller than that in the **L** direction, $j_{c,L}=j_{c,ab}$, flowing parallel to the *a-b* plane [3,11].

The magnetic field lines of superconducting films in a perpendicular magnetic field are strongly curved at small external fields. Even if we take into account that the vortices in a thin film will not generally exactly follow the magnetic field lines [19] it is nonetheless the case, except where $d \ll \lambda$, that the angle between the vortices and the cuprate planes will vary in different regions of the film. Different vortex orientations with respect to the crystal lattice directly lead to variations in $j_c$. Although a sharp change in the critical current is observed with the transition from a rectilinear FLL to a kinked FLL, the critical current anisotropy will also vary with θ above $θ_1$ since in an anisotropic Ginzburg-Landau superconductor the vortex cores are elliptical [20].

In the series of magneto-optical images shown in Fig. 6, the magnetic flux distribution after zero-field cooling is shown for different applied external fields. The rectangular shape of the sample determines the flux and current patterns (see Fig. 7), forming a current domain dominated by $j_{c,T}$ and two current domains dominated by $j_{c,L}$, separated by discontinuity lines where the current stream lines are sharply bent.

In the centre of the film, a discontinuity line separates two $j_{c,L}$ domains with different flow directions. In increasing field the discontinuity lines, separating **T** and **L** direction current flow, are visible as dark lines in the flux pattern (Fig. 6a-b), whereas in decreasing field they form bright lines which indicate a local maximum of $B_z$. For all magnetic fields the anisotropy of $j_{c,T}$ and $j_{c,L}$ can be easily seen by different penetration depths of the magnetic flux in the **L** and **T** directions. The anisotropy ratio $A_j = j_{c,L}/j_{c,T} = \arctan^{-1}(\alpha)$ can be determined from the angle α between the discontinuity line and the sample edge [21]. For a material which is isotropic in-plane α=45°. The anisotropy ratio $A_j$ changes gradually with the external magnetic field reflecting the different forms of $j_{c,T}(B)$ and $j_{c,L}(B)$, this is typical for superconductors with anisotropic current density and has also observed in vicinal films with a high density of APBs [8].

In addition to the in-plane critical current anisotropy visible in Figs 6 and 7 the critical current in the upper half of the film, $j_{c,LU}$, is not equal to the critical current in the lower half of the film, $j_{c,LB}$. One explanation for this is that the angle θ between the local flux and the cuprate planes is different for the areas of the film. This would lead to differing vortex core anisotropy and therefore different pinning forces on the vortices.

In order to investigate if the difference between $j_{c,LU}$ and $j_{c,LB}$ is indeed related to a variation of the angle, θ, of the vortices with respect to the *a-b* planes the magnetic field lines and θ in a cross section through the film were calculated from the measured current density distributions.

The magnetic field lines of the sample were calculated from the longitudinal current density $j_{c,L} = j_x$ by integrating the Biot Savart law for a strip with rectangular cross section $S$ with thickness $d$ and width $W$ numerically.

$$B_y = -\frac{\mu_0}{2\pi} \iint_S \frac{j_x(y',z')(z-z')}{(y-y')^2 + (z-z')^2} dy'dz'$$
$$B_z = \frac{\mu_0}{2\pi} \iint_S \frac{j_x(y',z')(y-y')}{(y-y')^2 + (z-z')^2} dy'dz'$$

(3)

Since it is not possible to determine the $z$ dependence of the current density by magneto-optical measurements at fixed measurement height, a $z$-independent $j_{c,L}$ was assumed. This might appear to be a serious limitation of this technique. However, since $d$ is not much larger than the London penetration depth $\lambda_{ab}$, the spatial variation of $j_{c,L}$ along $z$ should be smooth and the calculated anisotropy of the field angles in both current domains remains qualitatively correct.

Fig. 8 shows the magnetic field lines and the corresponding $j_{c,L}(y)$ for two cases with different curvatures: 8a and 8b after zero field cooling in an increasing $B_{ex}$ =32 mT, where the field lines have a convex curvature all over the cross section of the sample. For decreasing $B_{ex}$=48.8 mT the curvature is reversed for most parts of the cross section (see Fig.s 8c and 8d). The flux distributions for both cases are shown in Fig.s 6a and 6d, respectively.

An asymmetry of the angle of the field lines θ in both $j_{c,L}$ domains can be seen by examining the field line pattern in Fig. 8. Fig. 9 shows θ($y,z=d/2$) at the surface of the superconducting film for the two states depicted in Fig. 8. For increasing $B_{ex}$, θ($y$) is generally smaller in the $j_{c,LB}$ domain (y<0) than in the $j_{c,LU}$ domain, corresponding to the case $j_{c,LB} < j_{c,LU}$. For decreasing $B_{ex}$, this behaviour is reversed and the smaller θ(y) values are observed in the $j_{c,LU}$ domain (y>0), corresponding to the case $j_{c,LB} > j_{c,LU}$.

Apart from a small region near to the flux front for increasing $B_{ex}$, all values of θ are larger than $θ_1$, where the cross over between the rectilinear vortex state to a kinked vortex state takes place. In this vortex regime the anisotropy of the elliptical vortex cores will vary with θ. This results in a changing anisotropy of the pinning force since this is partly dependent on the length of the vortex core [4]. As can be seen from Fig. 8 and 9 the value of θ is different for the two sides of the film, this accounts for the variation in the observed local critical current.

Similar effects have not been observed in APB-rich films on vicinal substrates, where the anisotropic microstructure dominates the transport properties.

CONCLUSION

$YBa_2Cu_3O_{7-\delta}$ films were grown on as-supplied substrates and film growth was observed to proceed with any stacking faults or anti-phase boundaries that form at the substrate superconductor interface being healed within a few unit cells. In this paper we have shown that in vicinal $YBa_2Cu_3O_{7-\delta}$ films without APBs the intrinsic vortex channeling effect is observed. Furthermore the MO studies show anisotropic flux penetration caused by a variation in the critical current across the surface of the film. It is therefore clear that $YBa_2Cu_3O_{7-\delta}$ films grown on vicinal substrates without significant APB formation have quite different critical current properties to those with a high density of APB defects.

In MO measurements flux will preferentially penetrate along APB defects if present, leading to the filamentary flux penetration previously observed on APB rich vicinal films. This is not the case in the films studied which show a smooth flux front. The MO measurements confirm the observation that vicinal $YBa_2Cu_3O_{7-\delta}$ films exhibit a significant in-plane critical current anisotropy due to the fact that currents flowing in the **T** direction must flow to some extent in the *c*-axis to cross between *a-b* planes. The difference in the critical currents in the **L** direction observed between the top and bottom of the films is due to the fact that the local angle between the flux vortices and the *a-b* planes is different. This local value of θ results (for θ>θ$_2$) in a variation in the shape of the vortex cores and thus a variation in the effectiveness of the pinning on the vortices. Although the MO technique is only sensitive to the magnetic field distribution at the surface of the film, it follows from this observation that the shape of the vortex cores and thus the pinning on the vortex lines may, for these small fields, vary through the thickness of the film. Moreover the effects seen would not occur if films of the thickness studied (220nm) has reached the thin film limit described by Brandt et al [19]. The transport measurements presented provide data for the angular region where the flux line lattice structure has changed to the kinked flux pancake/string structure. Due to the small fields employed in the MO study it was not possible to image this regime.

Comparing the properties of the low APB density films studied here to those of films with large APB densities discussed in earlier work it is apparent that APB defects are strong pinning centres which significantly increase $j_c$. The deliberate introduction of APB defects may well therefore be a route to enhancing the critical current in *c*-axis oriented films. Finally several techniques currently used for the production of biaxially textured coated conductors result in grains which are vicinally orientated, ensuring that such grains grow with a APB rich structure will be important to obtaining the highest possible in grain critical current.

Figure Captions, Durrell et al.

Fig. 1 The experimental geometry for current flowing in a track crossing the cuprate planes. θ is the angle between the cuprate planes and the applied field, φ is the angle of rotation of the plane in which θ is swept. The dashed line corresponds to θ=0º φ=0º and the dotted line to θ=0º φ=90º.

Fig. 2 Schematic of anti-phase boundaries and stacking faults in YBCO

Fig. 3 Cross sectional HREM picture of the microstructure above the $SrTiO_3$ / YBCO interface showing the tilted *a-b* planes. The inset shows an anti phase boundary that is healed after a few unit cells.

Fig. 4 Variation of critical current density with field angle in a 4 degree mis-cut film

Fig.5 Variation of critical current density with substrate mis-cut angle in the films studied.

Fig. 6 Magneto-optically measured flux distributions of a YBCO film on 6° tilted vicinal $SrTiO_3$ after zero-field cooling to 4.2 K. The tilt direction is parallel to the short samples edge.  a)  partly penetrated state with $B_{ex}$ = 32 mT, b) fully penetrated state $B_{ex}$ =  $B_{max}$ = 120 mT, c) $B_{ex}$ is decreased to 96 mT, d) $B_{ex}$ = 48.8 mT, e)  $B_{ex}$ = 40.8 mT and f) remanent state with overlay showing the discontinuity lines between flux domains and the angle α indicating the in-plane critical current anisotropy.

Fig. 7 Flux pattern of figure 6 d)  ($B_{ex}$=48.8 mT in decreasing field) together with the current stream lines and the profiles of the current density components $j_{c,L}$ and $j_{c,T}$. The additional anisotropy of $j_{c,LB}$ and $j_{c,LU}$ is directly visible in the $j_{c,L}$ profile, where $j_{c,LB} > j_{c,LU}$. The two d⁻ lines separate a region, where the $j_{c,LB}$ domain of virgin flux penetration in increasing field is still preserved (see arrow in the $j_{c,L}$ profile).

Fig. 8 Magnetic field lines and the corresponding $j_{c,L}(y)$ for two cases with different curvatures: 8a and 8b after ZFC in an increasing $B_{ex}$ = 32 mT, where the field lines have a convex curvature all over the cross section of the sample. For decreasing $B_{ex}$=48.8 mT the curvature is reversed at most part of the cross section (see Figs 8c and 8d). The flux distributions for both case are shown in Figs 6a and 6d, respectively.

Fig. 9 Local angle between flux vortices and the *a-b* planes across the middle of the film at the film surface, $z=d/2$, for increasing and decreasing external fields as used in Fig. 8.

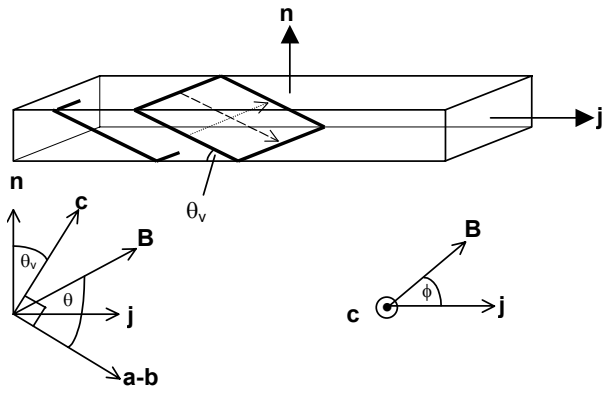

Figure 1, Durrell et al.

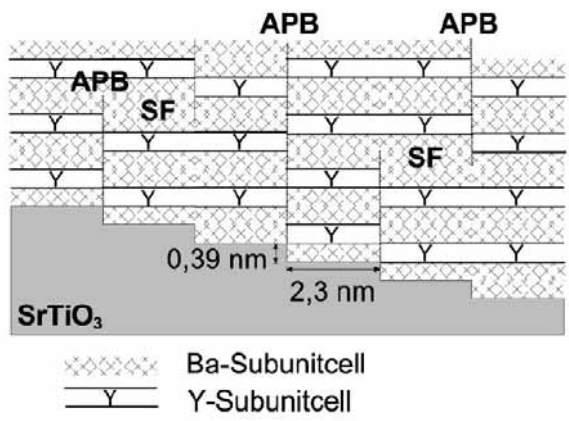

Figure 2, Durrell et al

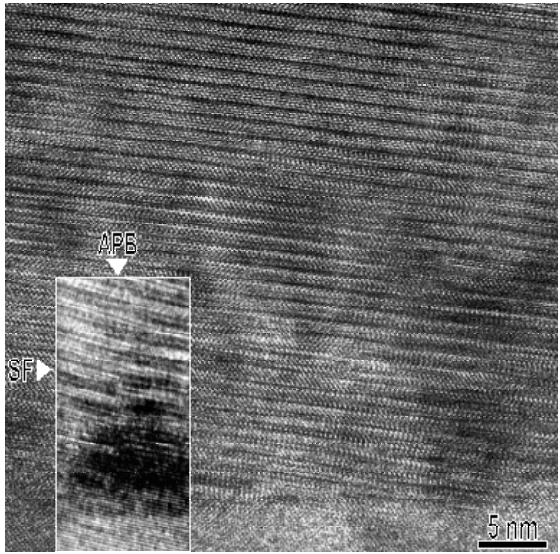

Figure 3, Durrell et al

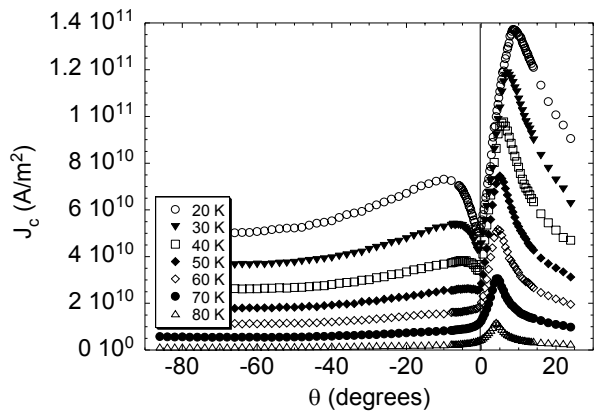

Figure 4, Durrell et al.

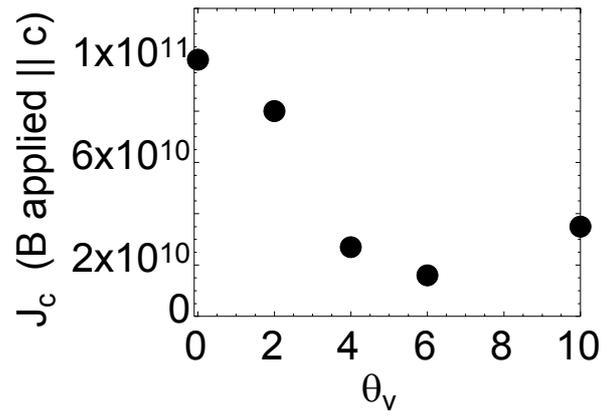

Figure 5, Durrell et al.

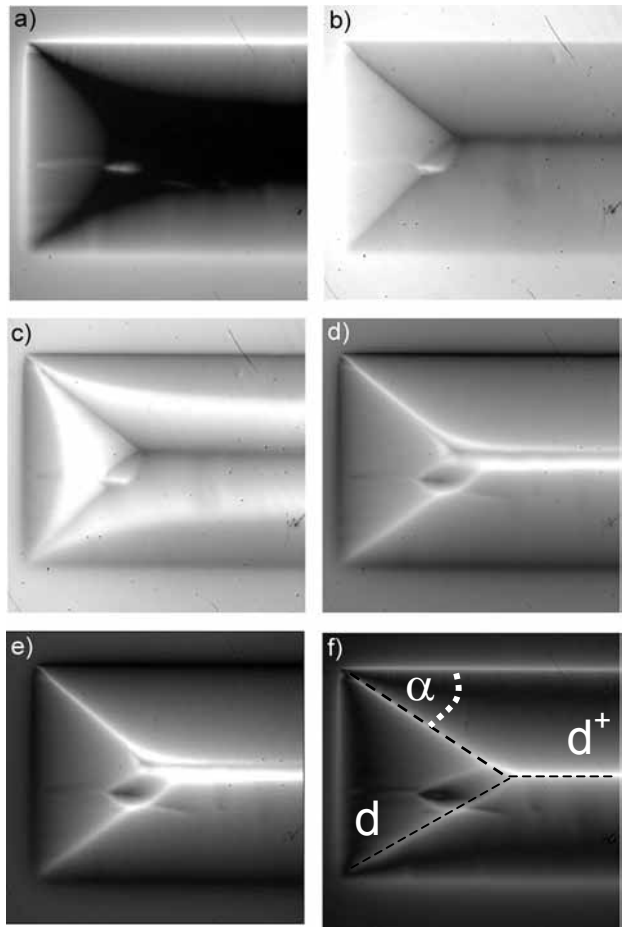

Figure 6, Durrell et al.

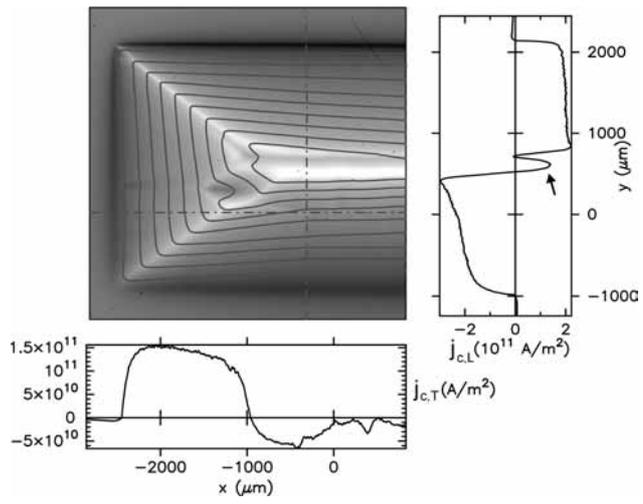

Figure 7, Durrell et al.

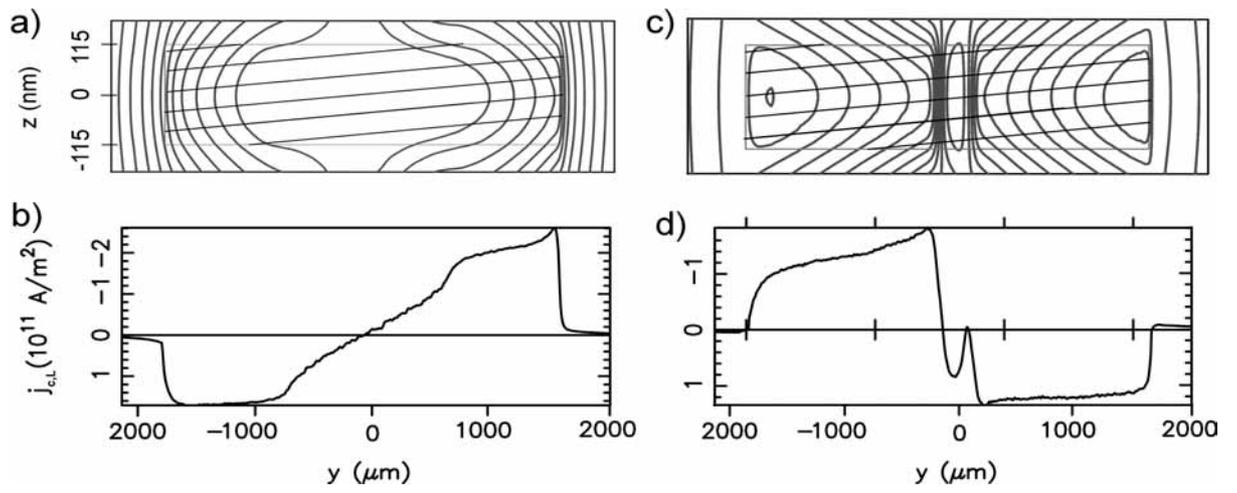

Figure 8, Durrell et al. (Wide)

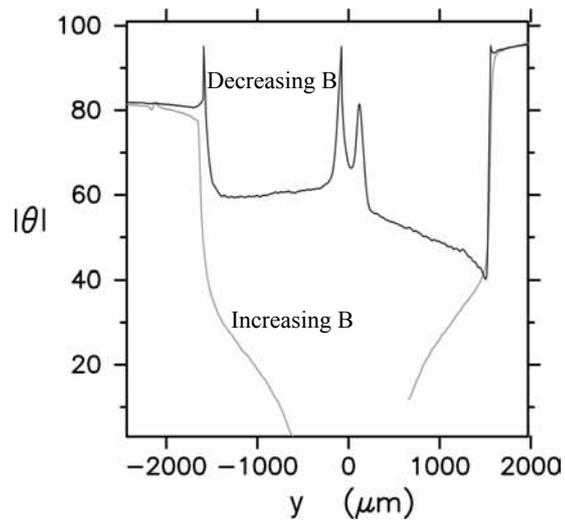

Figure 9, Durrell et al.

REFERENCES, Durrell et al.